\documentclass{svproc}

\usepackage{url}

\usepackage{overpic}
\newcommand{\piclabD}[2]{
\begin{overpic}[width=0.45\linewidth]{#1}
 \put (3,98) {\large \textsf{#2}}
\end{overpic}
}

\usepackage{overpic}

\usepackage{graphicx}
\usepackage{hyperref}
\hypersetup{
    colorlinks=true,
    linkcolor=blue,
    filecolor=blue, 
    urlcolor=blue, 
}

\begin{document}
\mainmatter              

\title{Ensemble approach for generalized network dismantling}

\titlerunning{Ensemble approach for generalized network dismantling}  
%
\author{Xiao-Long Ren \and Nino Antulov-Fantulin}
\authorrunning{Xiao-Long Ren \and Nino Antulov-Fantulin} 
%
\tocauthor{Xiao-Long Ren, and Nino Antulov-Fantulin}
\institute{ETH Zurich, Zürich CH-8092, Switzerland \\
To whom correspondence should be addressed: 
\email{ anino@ethz.ch}}

\maketitle              

\begin{abstract}
Finding a set of nodes in a network, whose removal fragments the network below some target size at minimal cost is called network dismantling problem and it belongs to the NP-hard computational class. 
In this paper, we explore the (generalized) network dismantling problem by exploring the spectral approximation with the variant of the power-iteration method.
In particular, we explore the network dismantling solution landscape by creating the ensemble of possible solutions from different initial conditions 
and a different number of iterations of the spectral approximation. 
\keywords{network dismantling, spectral partitioning, robustness}
\end{abstract}
\section{Introduction}
The process of network (graph) fragmentation by removing nodes or edges has a long history \cite{fiedler73,Lipton1979,Pothen1990,Bui1992,Guattery1998,NcutAttackArxiv} due to its practical relevance for maintaining the robustness of real-world systems \cite{Albert2000,Cohen2001,SchneiderPNAS,Gallos2006AttackStrategies,Qin2019}, containing contagion processes \cite{Pastor2002,Pastor2015,AntulovFantulin2015} or identification of node importance \cite{Lu2016}.
In case of vertex or edge separators, we want to find a small separator $S$ whose removal results in the partition to two roughly equal size \cite{Marx2006} sets. Finding the minimum vertex or edge separator for general graphs is an NP-hard problem \cite{Bui1992}, and it was approximated by different methods of 
linear programming \cite{Leighton1999}, semidefinite programming \cite{Arora2004,Feige2008,Arora2005SDP}, and spectral partitioning \cite{fiedler73,Pothen1990,Guattery1998}. 


In this paper, we will study the network dismantling problem \cite{K-separator,janson2008,Wu2016,Wu2018,ShlomoRobustnessAttacks}. A set $S$ is called a $C$-dismantling set if the largest/giant connected component (GCC) of a network contains at most $C$ nodes after removing the nodes in set $S$ \cite{K-separator,janson2008}.
Finding a minimum $C$-dismantling set is called network dismantling problem \cite{Braunstein2016}. Similarly, for a given network $G(V,E)$ with nodal costs $W=(w_1,\ldots,w_{|V|})$, the generalized network dismantling \cite{GND-PNAS} aims to find a set of nodes $S(G,W,C) \subseteq  V$ with the minimum dismantling cost, which will result in a fragmentation of the network into components of size at most $C$. 
The network dismantling problem belongs to the NP-hard class \cite{Braunstein2016} and can not have a fully polynomial-time approximation scheme (FPTAS) \cite{GND-PNAS} because of the theorem about the hardness approximating minimum vertex cover \cite{Dinur04onthe}.
This has motivated us to explore the ensemble of the dismantling solutions, instead of focusing on the (over)optimization within the vicinity of one potential point in the optimization function. 
Furthermore, our approach is also motivated by the studies of analyzing energy landscapes of highly non-linear optimization (loss) functions in machine learning \cite{LandscapeML} and modularity maximization in network science \cite{landscapeExploring,ClausetModularity}. In this paper, the exploration of dismantling landscape will be done by different initial conditions in the spectral partitioning method of generalized network dismantling method \cite{GND-PNAS}.

\section{Network dismantling approaches}

To solve the network dismantling problems, many efforts have been devoted recently. We will briefly introduce several representative algorithms we compared in this paper below.

Mugisha and Zhou \cite{BPAttack} related the network dismantling problem to the feedback vertex set problem and applied the belief propagation-guided decimation
(\textbf{BPD}) algorithm \cite{Zhou2013FVS} to solve it. BPD algorithm is a loop-focused global algorithm which removes the nodes with the highest probability to break most loops in the network. 

Braunsteina \textit{et al.} \cite{Braunstein2016} introduced a very efficient algorithm, \textbf{Min-Sum}, which consisted of three stages: (1) Using Min-Sum message passing to break all the loops in the network, then only trees are left. (2) Breaking all the trees whose size are bigger than the target dismantling size (threshold). (3) Greedily reinsert \cite{morone2015influence} the removed nodes that had been removed from the network in the previous two stages.  

\begin{figure}
\centering
\includegraphics[width=\linewidth]{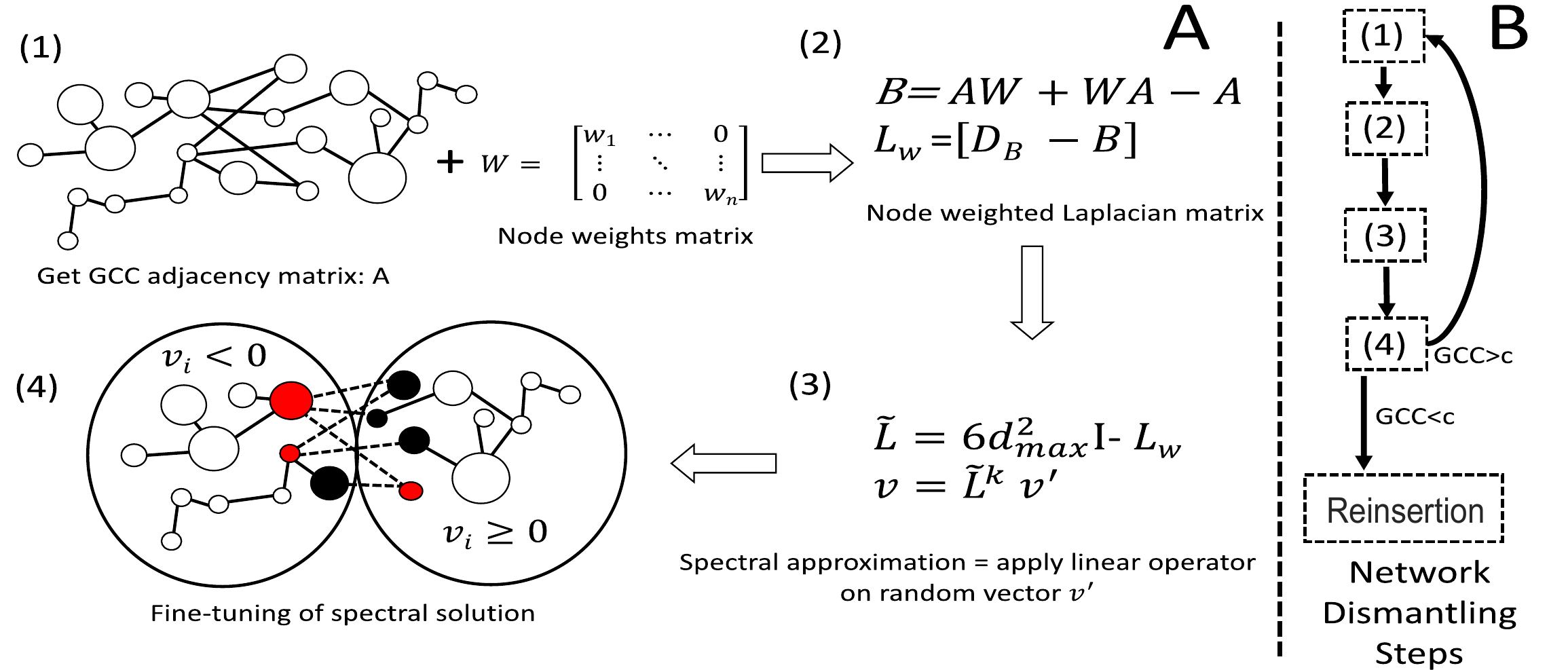}
\caption{A brief explanation of the procedure of the standard GND algorithm, see more details in ref. \cite{GND-PNAS}. When the size of the gaint connected components (GCC) is smaller than the target size, the algorithm will stop to remove more nodes and start to reinsert the removed nodes.
This figure comes from ref. \cite{GND-PNAS}.}
\label{fig:GND}
\end{figure}

More recently, Ren \textit{et al.} \cite{GND-PNAS} studied the generalized network dismantling problem, which aims at finding a set of nodes with minimal dismantling cost. The dismantling cost can be any arbitrary non-negative real values. To solve this expanded problem, they proposed the \textbf{GND} algorithm
which is based on the iterative node-weighted spectral approximation and a fine-tuning weighted vertex cover method \cite{BARYEHUDA1981}. Please find more details in Fig. \ref{fig:GND} and Section \ref{section:Ensemble-GND}.

Fan \textit{et al.} \cite{Fan2019} reformulated the network dismantling problem as a Markov decision process and employed deep reinforcement learning to train the \textbf{GraphDQN} agent 
to efficiently solve the problem. To the best of our knowledge, this is the first practice to solve the (generalized) network dismantling problem by using deep
reinforcement learning approach. 

In addition to the algorithms introduced above, there are also many other commonly used algorithms, such as equal graph partitioning (EGP) \cite{Chen2008EGP}, Collective Influence (CI) \cite{morone2015influence}, CoreHD \cite{Zdeborova2016}, and so on \cite{Buluc2016}. Detailed comparisons of these algorithms can be found in ref. \cite{GND-PNAS,Fan2019}. 

\section{Ensemble-GND algorithm}
\label{section:Ensemble-GND}
\subsubsection{More iterations or more random tries?}
\label{subsection:Moreiterations}

The detailed procedure of the standard GND algorithm was elaborated in the paper \cite{GND-PNAS}. The standard GND algorithm uses a variant of the power-iteration method to calculate the eigenvector of the second smallest eigenvalue of the weighted Laplacian matrix of the network. The spectral approximation uses the deterministic initialization with the pseudorandom Mersenne Twister generator \cite{Matsumoto} with default seed.
The authors showed that in every bisection, it usually takes $P= O(log(n) * \sqrt{(log(n))} )$ iterations to get an effective eigenvector so that it can get a good partition. In particular, they have used $P= 30 * log(n) * \sqrt{(log(n))}$ iterations.

The following question arises: Is it possible to obtain better results by allowing more iterations in the power method?
Here we test the results when the number of iterations is $D*P$ (other conditions and parameters keep the same). The result on Petster-hamster network dataset \cite{KONECT} is shown below in Fig. \ref{fig:Comparison1000}. The blue curves in this figure are the standard/original GND(R) algorithms published in paper \cite{GND-PNAS}. The green curves are the results of the tested procedure with $D=1000$ times more iteration. This result suggests that more iterations in the power method doesn't necessary produce better dismantling result in the GND algorithm.

\begin{figure}
\centering
\includegraphics[width=0.6 \textwidth]{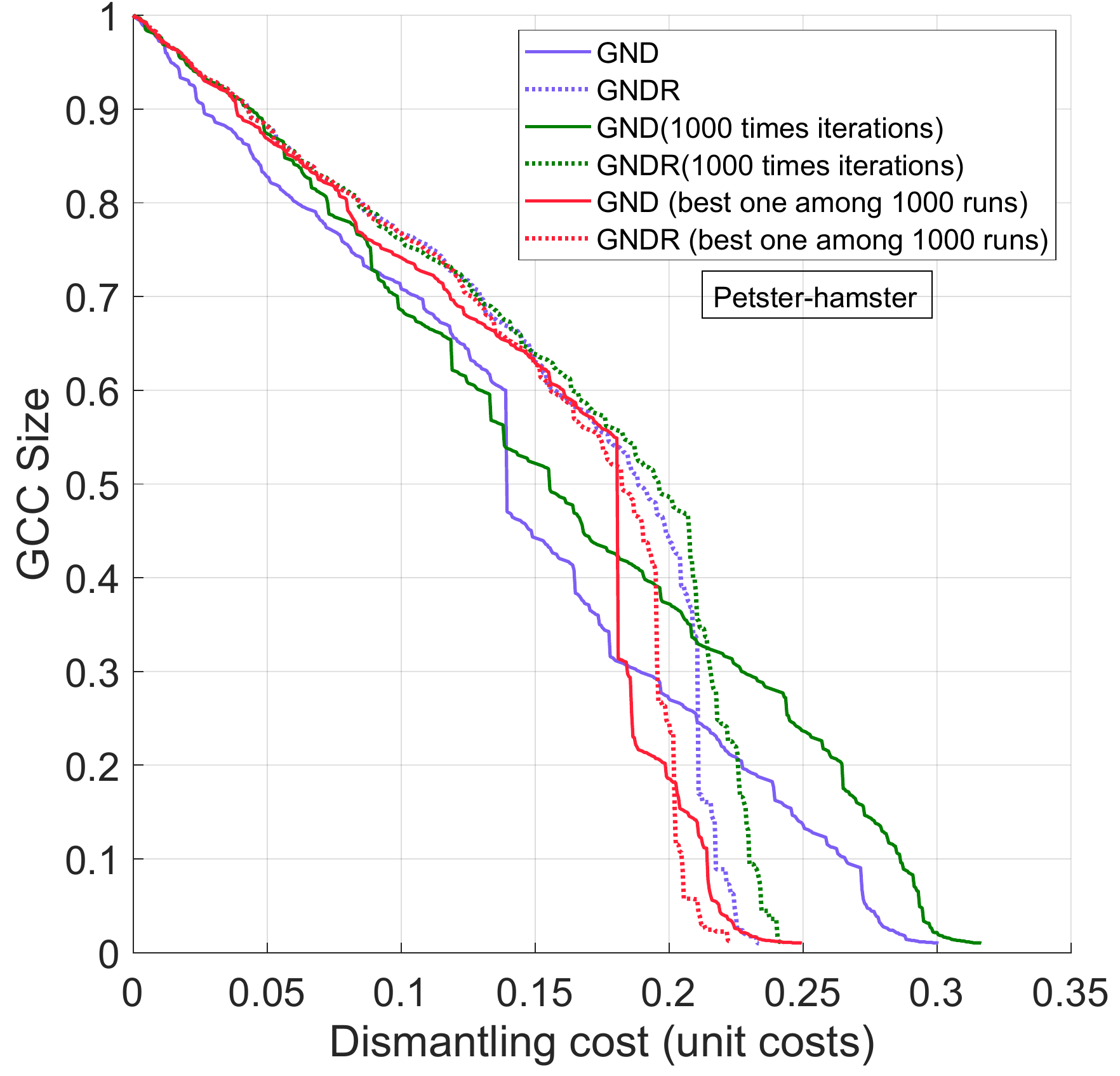}
\caption{Comparison of the standard GND(R) algorithms with its variants. The blue curves are the standard GND and GNDR algorithms published in ref. \cite{GND-PNAS}. 
The green curves are the standard GND(R) algorithms with $D=1000$ times more iterations in the power method when computing the eigenvector in every bisection.}
\label{fig:Comparison1000}
\end{figure}

It is not surprising that being more accurate in approximating eigenvectors does not lead to better dismantling solution. After all, the problem is NP-hard, and the spectral formulation is just the integer relaxation \cite{Guattery1998} without strict bounds on the optimality. 
To contrast, we also tested another approach. As we know, every time before calculating the eigenvector in every bisection, the GND algorithm need to produce an initial vector $v$. Then after $P=30 * log(n) * \sqrt{(log(n))}$ iterations the power method will get the approximation of the eigenvector. According to this approximation, all the nodes in the GCC of the network will be partition into two parts, $M$ and $\bar{M}$. Then the weighted vertex cover method will be applied and GND can get the set of nodes that should be removed in this bisection. 
As we can see, the result of the bisection is based on the the initial vector $v$ which is always deterministic, due to the default seed of the pseudorandom number generator. Default seed is used for the reproducibility purposes. 

\begin{figure}[!htb]
\begin{center}
\piclabD{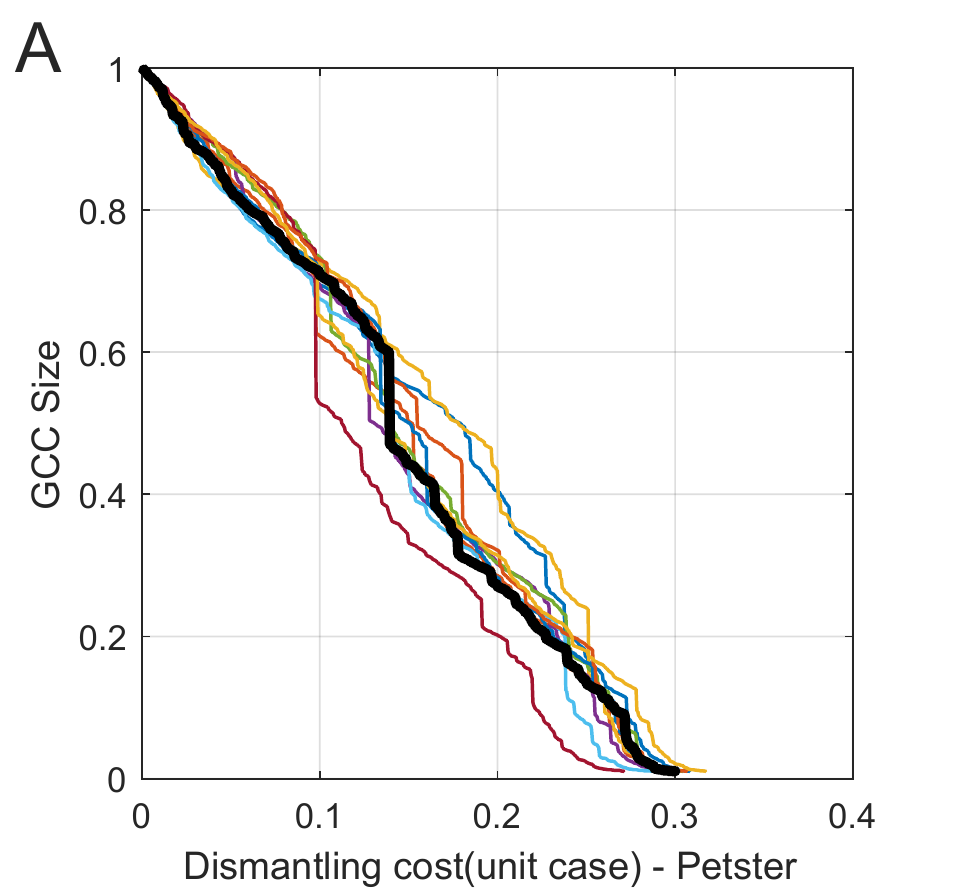}{}
\piclabD{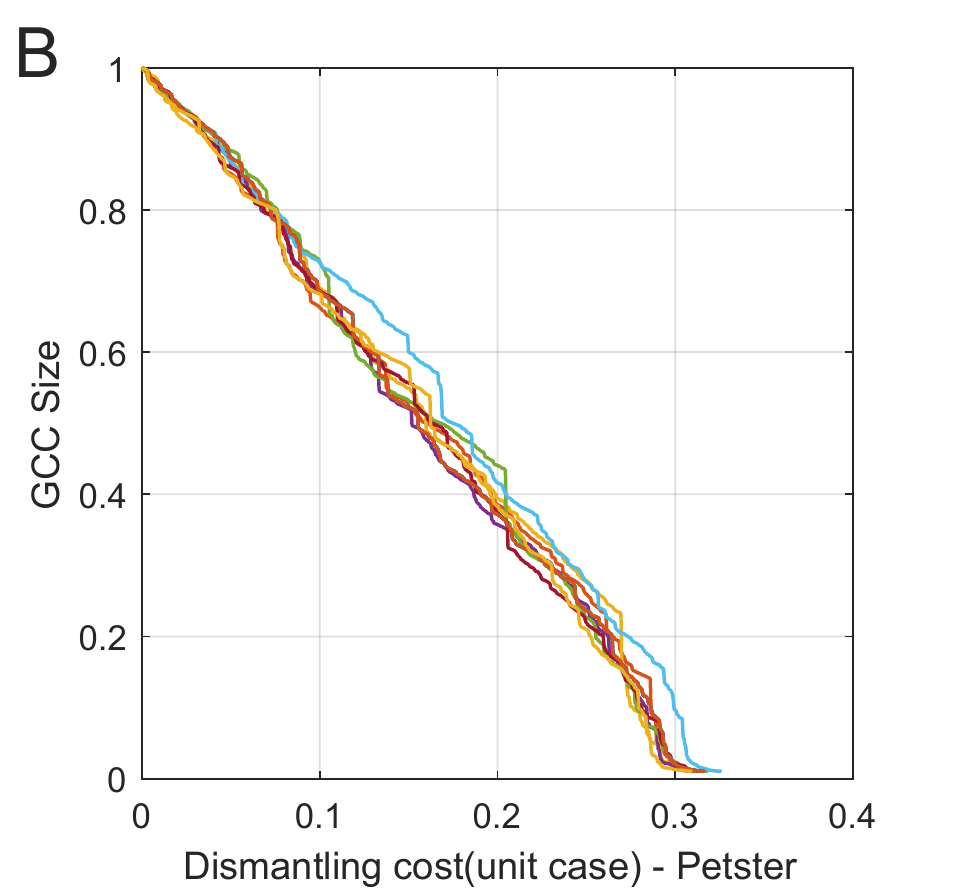}{}
\piclabD{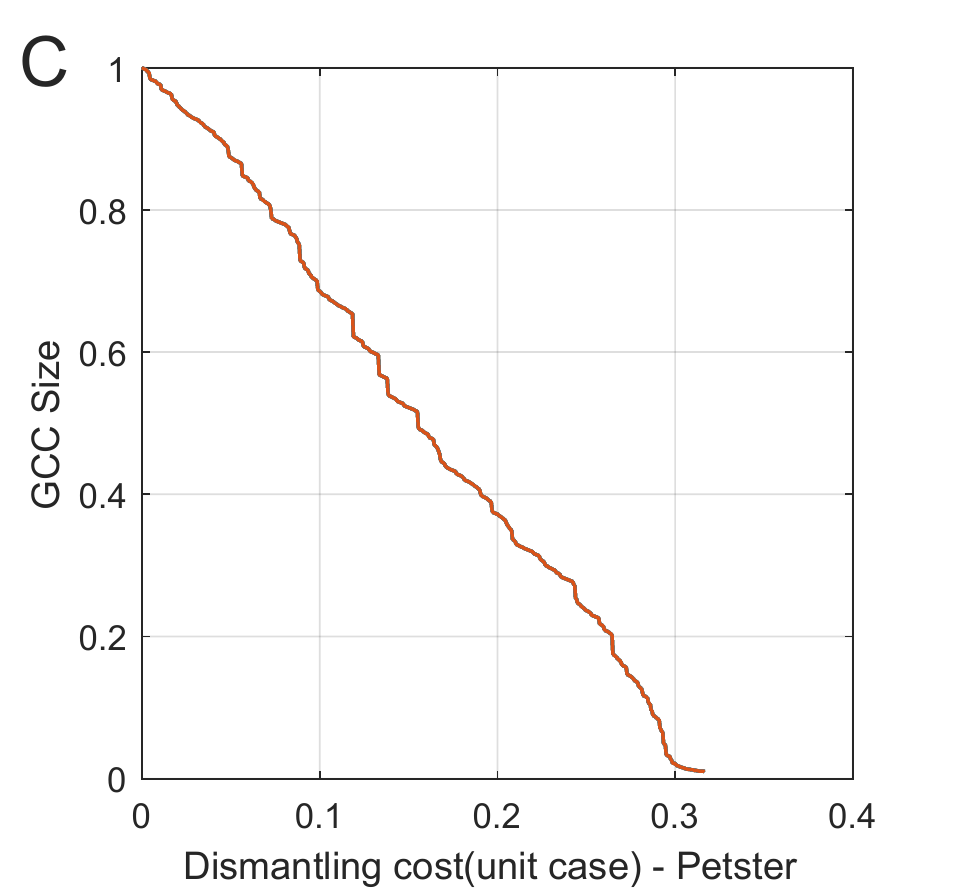}{}
\piclabD{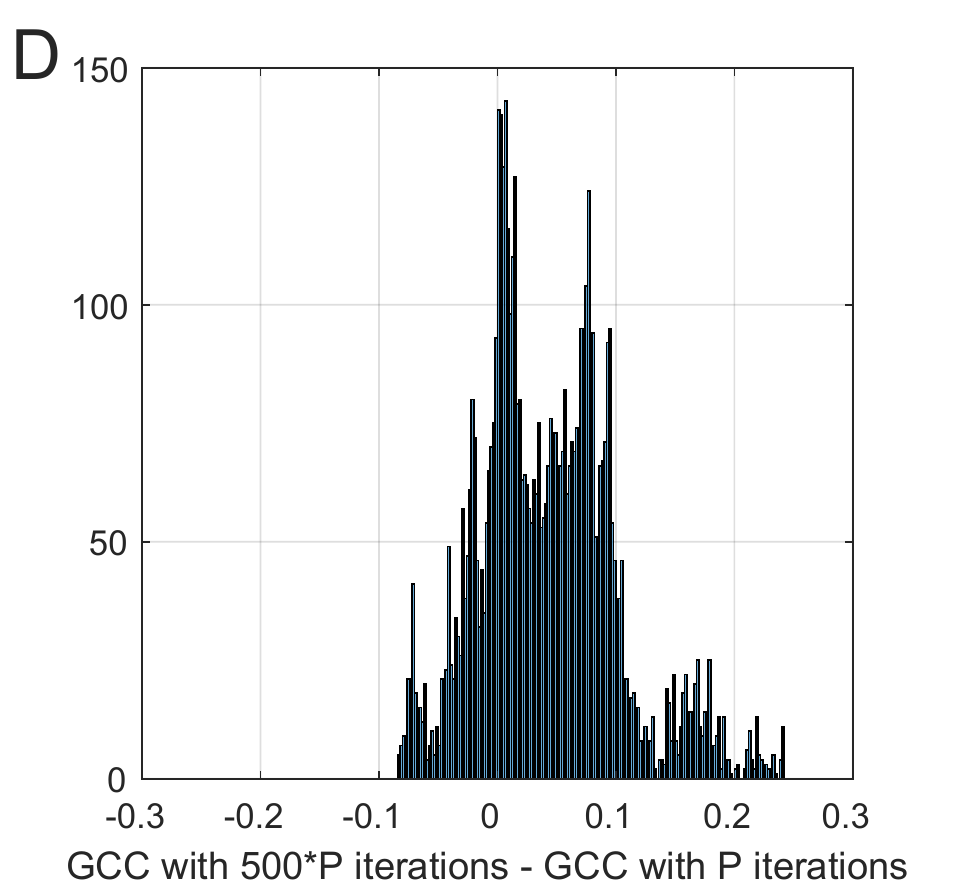}{}
\end{center}
\caption{{\bf Variability of the ensemble approach of GND.} (A)-(C) show numerical results for 10 different initializations: (A) with $P$ iterations for each initialization, (B) with $200*P$ iterations for each initialization, (C) with $500*P$ iterations for each initialization, where $P=30 * \log(n) * \sqrt{log(n)}$. When number of iterations $D*P$ exceeds $D=500$, we no longer see the variability from different initializations.
(D) We measure the difference in the dismantling performance $GCC_{500*P}(c)-GCC_{P}(c)$, where $GCC_{x}(c)$ denotes the GND algorithm with $x$ spectral approximation iterations for cost $c$.
The graphic shows the histogram of differences in GCC over all possible costs for different seeds. We observe that the majority of differences is positive, which implies having a smaller GCC for the same cost for setting with $P$ iterations.}
\label{fig:diff_inits}
\end{figure}

Alternative approach of using single initialization with $D*P$ iterations is to use $K$ different initialization with $P$ iterations, which we call ensemble approach. In order to have the same run-time complexity, we will fix $K=D$.
In the ensemble approach, we will produce $K$ different dismantling solutions $S_1$,$S_2$,...$S_K$, and take the one with the with the \textbf{minimum cost} 
\[
S^*=\min{ \lbrace S1, S2, ..., S_K \rbrace }.
\]
The statistical behavior of the minimum cost is given by the extreme value distribution, however, in this paper we only use deterministic approach. 
Different initializations are produced with the pseudorandom number generator with default seed, which results with deterministic method. 
This approach has the same computational complexity and similar running time with the method we tried above (green curves). The results of this approach are the red curves shown in Fig. \ref{fig:Comparison1000}. We can see that this variant GND and its GNDR algorithm (red curves) have much better performances than the standard GND and GNDR algorithms. In Fig. \ref{fig:diff_inits}, we show the variability of the ensemble approach that was explored with $K=10$ different initializations. 

\subsubsection{Fine-tuning of the initial partition in every bisection of the GND algorithm.}
\label{subsection:Fine-tuning}
In the standard GND algorithm, after getting the eigenvector corresponding to the second smallest eigenvalue of the weighted Laplacian matrix, all the nodes will be partition into two groups, $M$ and $\bar{M}$, according to their values in the eigenvector. More specifically, the nodes with a value smaller than $0$ will be put in group $M$, else group $\bar{M}$. After this, the links between the two groups should be removed to partition the network into two disconnected parts (see Fig. \ref{fig:GND}\textcolor{blue}{A(4)} and ref. \cite{GND-PNAS}). Then the weighted vertex cover method will be applied to fine-tuning the dismantling set.

\begin{figure}
\centering
\includegraphics[width=0.9\textwidth]{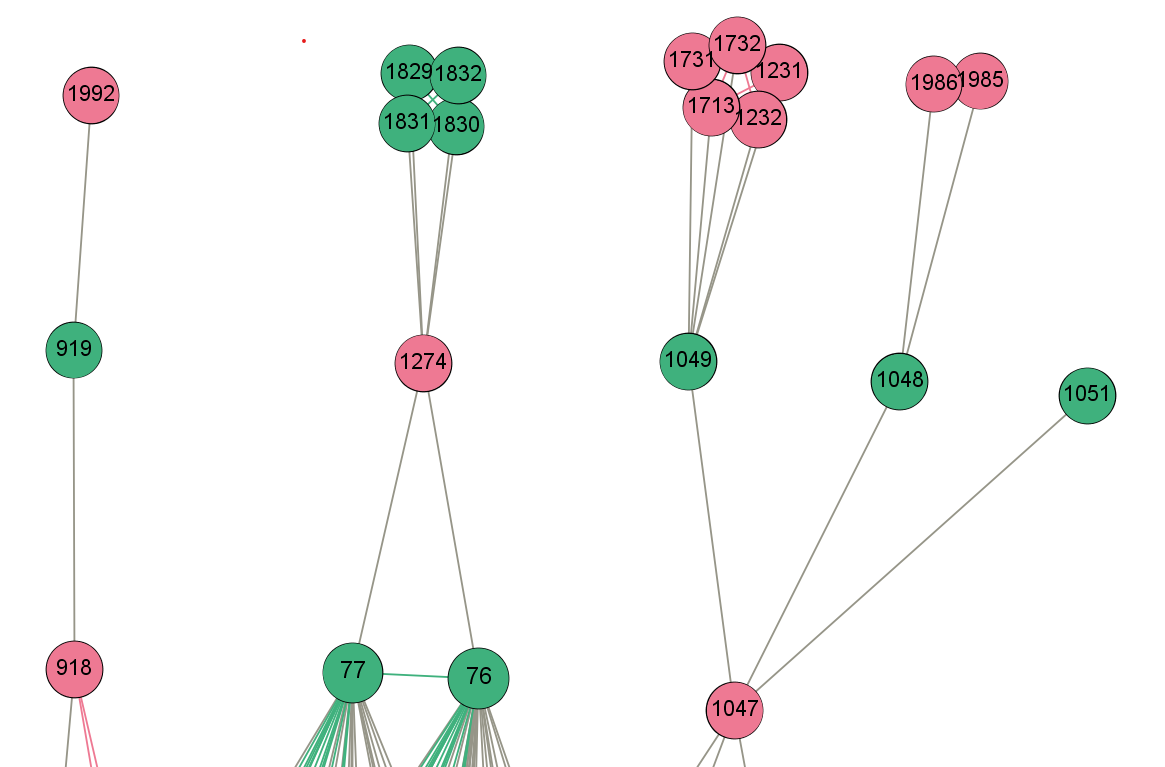}
\caption{An example of partitioning the nodes in Petster-hamster network \cite{KONECT} into two groups according to their values in the eigenvector. The green nodes have values smaller than $0$ (group $M$) while the red nodes have values equal to or bigger than $0$ (group $\bar{M}$).}
\label{fig:partition_petster}
\end{figure}

However, the initial partition of the two groups exactly according to nodes' value in its eigenvector is not always a perfect choice. We exemplify this fact in Fig. \ref{fig:partition_petster}, which shows an instance of partitioning the nodes in Petster-hamster network into two groups according to their values in its eigenvector. All the green nodes have values smaller than $0$ (group $M$) while all the red nodes have values equal to or bigger than $0$ (group $\bar{M}$). Thus the edges between the two groups should be removed to make the group disconnected. Then the vertex cover method will be employed to find the fine-tuning dismantling set of nodes based on this step.
Please note that there are $N=2,000$ nodes in the Petster-hamster network and the target size of the dismantling is $C=1\%*N=20$. In this example, removing nodes $\{919, 1274, 1049, 1048, 1051\}$ does not reduce the size of GCC below the target but is increasing the dismantling cost. 
To improve the performance of the standard GND algorithm, the partition of the nodes should be adjusted in the above examples. Thus we will adjust the ascription (i.e., $M$ or $\bar{M}$) of a node if all of its neighbors are belonging to another group. 
For example, the nodes $\{919, 1049, 1048, 1051\}$ will be adjusted to red while the node $\{1274\}$ will become green after adjustment and all these nodes do not need to be removed.
The \textbf{spectral fine-tuning} can be formalized as the following rule. For an arbitrary node $v$ belonging to group $M$ with cardinality $|M|>1$, if all its neighbors belong to the opposite group $\bar{M}$, we change $v$'s group label to the $\bar{M}$. Intuitively, this rule \textbf{filters the noise} from numerical approximation of spectral partitioning.

\subsubsection{The Ensemble-GND and Ensemble-GNDR algorithm.}
We propose a variant of the standard GND(R) algorithm called \textbf{Ensemble-GND} and \textbf{Ensemble-GNDR} which consider the upper two issues in this section, that is, based on the standard GND algorithm \cite{GND-PNAS}, the Ensemble-GND(R) algorithm will (1) adjust the partition of the nodes in group $M$ and $\bar{M}$ according to their surrounding connectivity for specific target size (see details from the previous paragraphs of this section), and (2) select the result with the best performance in the ensemble with $D$ ($D=1000$ in this paper) results produced with $D$ different initializations (instead of one initialization with the default seed, see details from the previous parts of this section).

\begin{table}[]
\caption{Properties of the networks we used in this paper. }
\label{table:NetProperties}
\centering
\begin{tabular}{lccccccc}
\hline
  & Crime  & Petster-hamster  &  RoadEU & Political-blogs & Crime2 & HI-II-14 & DBLP \\ \hline
Nodes & 754 & 2000  &  1177 & 1222  & 829  & 4165  & 12495 \\
Links & 2127 & 16714  & 1305 & 16714 & 1473  & 13087  & 49563 \\
\hline
\end{tabular}
\end{table}

In this paper we use seven popular real world network datasets \cite{KONECT,Zdeborova2016,Fan2019} to compare the performance of the existing algorithms and the proposed Ensemble-GND and Ensemble-GNDR algorithms. The properties of the networks are listed in Table \ref{table:NetProperties}. For all the dismantling tasks, we set the dismantling target size as the 1\% of the original network size, i.e., when the GCC of the remaining network is smaller than the target size, the algorithm will stop to remove more nodes. In addition, for the unweighted cost case, the dismantling cost of any arbitrary nodes is the same. For the weighted cost case, the removal cost of any arbitrary nodes is equal to its remaining degree in the network. The running time of the one initialization of the GND algorithm are summarized in Table \ref{table:runningtime}.

\begin{table}[]
\caption{The running time for one round of Ensemble-GND algorithm.}
\label{table:runningtime}
\centering
\begin{tabular}{lccccccc}
\hline
 Running time(s) & Crime  & Petster-hamster  &  RoadEU & Political-blogs & Crime2 & HI-II-14 & DBLP \\ \hline
Unweighted  & 0.147 & 1.200  &  0.193 & 1.184  & 0.146  & 1.047  & 23.551  \\
Weighted case & 0.407 & 3.798  & 0.374 & 4.151 & 0.426  & 4.730  &  53.399 \\

\hline
\end{tabular}
\end{table}

The results of the Ensemble-GND(R) algorithm are listed in Table \ref{table:compareGND} and Table \ref{table:compareGraphDQN}. In Table \ref{table:compareGND}, we compared our Ensemble-GND(R) algorithm with the state-of-the-art algorithms, including BPD, Min-Sum, and GND(R) algorithms, for weighted dismantling cost case and uniform cost (unweighted) case, respectively. We can clearly see that for all the weighted case and almost all the unweighted case (except the Political-blogs network), the Ensemble-GNDR can obtain the best performance. 

\begin{table}[]
\caption{Comparison of the standard GND(R) and Ensemble-GND(R) algorithm by the dismantling cost. The better results are highlighted with bold text.}
\label{table:compareGND}
\centering
\begin{tabular}{lcccccc}
\hline
Unweighted case & BPD & Min-Sum &  GND  & Ensemble-GND  & GNDR  & Ensemble-GNDR \\ \hline
Crime Network & 101& 120 &  110 & {103} & 103 & \textbf{99} \\
Petster Network & 474 & 485 &   601& {510} & 467 & \textbf{441} \\ 
RoadEU Network & 151 & 160 &  193& {159} & 171 & \textbf{144} \\
Political-blogs & \textbf{375} & 380 &  494& 435 & 404 & 386 \\
\hline \hline
Weighted case & BPD & Min-Sum & GND  & Ensemble-GND  & GNDR  & Ensemble-GNDR \\ \hline
Crime Network & 0.594 & 0.644 & 0.642 &{ 0.624} & 0.584 &  \textbf{0.572} \\
Petster Network & 0.829 & 0.837 & 0.914 & {0.873} & 0.810 & \textbf{0.792} \\ 
RoadEU Network & 0.463 & 0.491 & 0.523 & {0.470} & 0.464 & \textbf{0.417} \\ 
Political-blogs & 0.978 & 0.979 & 0.995 & {0.993} & 0.984 & \textbf{0.977} \\ 
\hline
\end{tabular}
\end{table}

Further more, we also compared our algorithm with the deep reinforcement learning approach GraphDQN. 
The code of the GraphDQN method was not available at the time of writing of this paper.
Therefore, we have made comparisons only on part of the networks, for which we had GraphDQN dismantling solutions (provided by the authors of study \cite{Fan2019}).
Based on all the five results in Table \ref{table:compareGraphDQN}, we can conclude that the proposed Ensemble-GNDR algorithm has better performance than the deep reinforcement learning approach in both unweighted and weighted cost cases.

\begin{table}[]
\caption{Comparison of the Ensemble-GND(R) and the deep reinforcement learning-based algorithm GraphDQN. The dismantling results of the GraphDQN were obtained from the authors of the paper \cite{Fan2019} (The code of the GraphDQN algorithm was not available at time we were writing this paper, but only the solutions on several datasets that we have used). The better results are highlighted with bold text. }
\label{table:compareGraphDQN}
\centering
\begin{tabular}{lccc}
\hline
Unweighted case & GraphDQN  & Ensemble-GND  &  Ensemble-GNDR \\ \hline
Crime2 Network & 185 & 183  & \textbf{161} \\
HI-II-14 Network & 553 & 483  & \textbf{412} \\
DBLP Network &  2496& 2499 &  \textbf{2064} \\ 
\hline 
\hline
Weighted case & GraphDQN  & Ensemble-GND    & Ensemble-GNDR \\ \hline
Crime2 Network & 0.989 & 0.802 &  \textbf{0.718} \\
HI-II-14 Network & 0.977  & 0.942  & \textbf{0.831} \\
\hline
\end{tabular}
\end{table}


\section{Conclusion}
In this paper, we briefly reviewed the recent progress in the study of the network dismantling problem and explored the solution landscape of (generalized) network dismantling problem by proposing the Ensemble-GND and Ensemble-GNDR algorithm. 
We compared the proposed Ensemble-GND(R) algorithm with the state-of-the-art algorithms, including BPD, Min-Sum, and the standard GND(R) algorithms, as well as a recently proposed deep reinforcement learning-based algorithm GraphDQN. The results show that our Ensemble-GND(R) has a better performance both in the weighted case and the unweighted case of the network dismantling problem.
Which opens new research directions of exploring the ensemble of dismantling solutions in the objective landscape by different methods of perturbations, initializations, and other modern machine learning optimizations techniques for highly non-linear and non-convex objective functions. 

\section{Acknowledgements}
X.L.R. thanks to the financial support of China Scholarship Council (CSC).
N.A.-F. thanks to the financial support from the EU Horizon 2020 project SoBigData under grant agreement No. 654024.

%
%
\bibliographystyle{ieeetr}

\bibliography{main}

\end{document}